\documentclass[10pt,conference]{IEEEtran}

\IEEEoverridecommandlockouts

\usepackage[pdftex]{graphicx}
\graphicspath{{.}}
\DeclareGraphicsExtensions{.pdf,.jpeg,.png}

\makeatletter
\newcommand{\newlineauthors}{%
\end{@IEEEauthorhalign}\hfill\par
\hfill\begin{@IEEEauthorhalign}
}
\makeatother

\usepackage[T1]{fontenc} 
\usepackage{amsmath}
\usepackage[cmintegrals]{newtxmath}
\usepackage{bm} 

\usepackage[utf8]{inputenc}
\usepackage{float}
\usepackage{booktabs}
\usepackage[obeyspaces]{url}

\begin{document}
	\title{ODRE Workshop: Probabilistic Dynamic Hard Real-Time Scheduling in HPC %
		\thanks{This research has been funded iCyPhy, under Siemens sponsorship.}
		\thanks{\copyright 2020 IEEE. Personal use of this material is permitted. Permission from IEEE must be obtained for all other uses, in any current or future media, including reprinting/republishing this material for advertising or promotional purposes, creating new collective works, for resale or redistribution to servers or lists, or reuse of any copyrighted component of this work in other works.}
	}
	
	\markboth{11th {IEEE} International Conference on Cloud Computing Technology and Science (CloudCom 2019)}%
	{Hofer \MakeLowercase{\textit{et al.}}: Probabilistic dynamic hard real-time scheduling in HPC}
	
	\author{\IEEEauthorblockN{Florian Hofer,~\IEEEmembership{Student Member, IEEE}}
			\IEEEauthorblockA{\textit{Faculty of Computer Science} \\
				\textit{Free University of Bolzano-Bozen}\\
				Bolzano, Italy \\
				florian.hofer@stud-inf.unibz.it}
			\and
			\IEEEauthorblockN{Martin A. Sehr}
			\IEEEauthorblockA{\textit{Corporate Technology} \\
				\textit{Siemens Corporation}\\
				Berkeley, CA 94704, USA\\
				martin.sehr@siemens.com}
			\and 
			\IEEEauthorblockN{Barbara Russo,~\IEEEmembership{Member, IEEE}}
			\IEEEauthorblockA{\textit{Faculty of Computer Science} \\
				\textit{Free University of Bolzano-Bozen}\\
				Bolzano, Italy \\
				barbara.russo@unibz.it}

			\and
			\newlineauthors
			\IEEEauthorblockN{Alberto Sangiovanni-Vincentelli,~\IEEEmembership{Fellow, IEEE}}
			\IEEEauthorblockA{\textit{EECS Department} \\
				\textit{University of California}\\
				Berkeley, CA 94720, USA\\
				alberto@berkeley.edu}
			
		}

	\maketitle
	
	

	\begin{abstract}
		Industry 4.0 is changing fundamentally the way data is collected, stored and analyzed in industrial processes.
		While this change enables novel application such as flexible manufacturing of highly customized products, the 	
		real-time control of these processes, however, has not yet realized its full potential.
		We believe that modern virtualization techniques, specifically \textit{application containers}, present a unique opportunity to decouple control functionality from associated hardware.
		Through it, we can fully realize the potential for highly distributed and transferable industrial processes even with real-time constraints arising from time-critical sub-processes.
		In this paper, we present a specifically developed orchestration tool to manage the challenges and opportunities of shifting industrial control software from dedicated hardware to bare-metal servers or (edge) cloud computing platforms.
		Using \textit{off-the-shelf} technology, the proposed tool can manage the execution of containerized applications on shared resources without compromising hard real-time execution determinism.
		Through first experimental results, we confirm the viability and analyzed the behavior of resource shared systems with strict real-time requirements.
		We then describe experiments set out to deliver expected results and gather performance, application scope and limits of the presented approach.
	\end{abstract}
	
	\section{Introduction}

	Industry 4.0 is the outcome of a strategy-project of the German government, intended to increase the computerization of manufacturing.
	The major technological pillars of Industry 4.0 are Cyber Physical System (CPS), Internet of Things (IoT), Information and Communications Technology (ICT), Enterprise Architecture (EA), and Enterprise Integration (EI) \cite{Lu2017}.
	The four Industry 4.0 design principles \cite{Hermannetal2016}, (interconnection, information transparency, decentralized decisions and technical assistance), require high levels of interconnection and interdependence.
	As a result, the modular structured units perform monitoring, control and run virtual copies to facilitate decentralized decisions.
	Concepts like adaptive manufacturing systems and re-configurable production require to connect all levels of operation.
	Consequently, Edge Computing including cross-platform operation, third party software and mixed criticality applications are taking the central stage.
	The computation requirements given the migration of functionality towards the ``edge'' imply  new system architectures~\cite{Telschigetal2018, Hofer2018}. 
	These requirements range from data processing and performing at low criticality to hard-real-time control applications. 
	The allocation of resources on an edge node will thus impact the performance, including response time, and installation cost.

	We believe that modern virtualization techniques such as application containerization~\cite{Mogaetal2016,Tascietal2018,Goldschmidtetal2018} are essential for appropriate utilization of cloud computing resources.
	Industrial control systems can yield through such techniques the same advantages that traditional containerized micro-services present: the creation of light and easily distributed control applications able to run on any system and that are, at the same time, easy to maintain and update~\cite{Fazioetal2016}.
	Many legacy systems still exist that rely on now obsolete development tools but are still vital to plant function~\cite{HoferRusso2019}. 
	Virtualization of such systems could extend their life beyond previously defined limits, guaranteeing continuity of operation while extending features and function~\cite{GoldschmidtHauck-Stattelmann2016}.
	
	Current trends in industry favor the flexibility of resource sharing through cloud computing, but merging it with real-time requirements is a challenging task.
	Garcia-Vallas et al.~\cite{Garcia-Vallsetal2014} state that guest OSs have only limited access to physical hardware and thus suffer from unpredictability of non-hierarchical scheduling and thick-stack communications.
	Abeni et al.~\cite{Abenietal2018} extended the standard Completely Fair Scheduler hierarchically with a deadline based algorithm, optimizing latency results for containerized software.
	The modified scheduler successfully manages also a greater amount of time critical tasks, performing better than the default deadline based scheduler.
	While there exist real-time enabled hypervisors with direct access to hardware, such as the paravirtualized RT-Xen, the shared resources still exhibit latency that may make real-time execution difficult.

	Containerizing control applications as an alternative to traditional virtualization has also been addressed in the literature: 
	Moga et al.~\cite{Mogaetal2016}, for instance, presented the concept of containerization for full control applications to decouple hardware and software life-cycles of industrial automation systems. 
	Telschig et al.~\cite{Telschigetal2018} explore a platform-independent container architecture for real-time systems; with a dedicated architecture and prototype agent that manages communication between dependent distributed software, the authors focus on isolation of critical from non-critical tasks and their respective portability.
	Tasci et al.~\cite{Tascietal2018} presented a Linux-based solution as host operating system, including both the single kernel, preemption-focused PREEMPT-RT patch, and the co-kernel oriented Xenomai, where they evaluated both migration feasibility and performance. 
	However, these approaches ground on experimental research and do not consider resource optimization using Commercial-Off-The-Shelf (COTS) components.
	
	In this paper, we propose a technique to optimize the resource sharing on computation infrastructure.
	Using COTS technology, we assess the viability of our probabilistic resource sharing approach for the reduction of operating costs.
	
	\section{Resource management}
	\label{sec:resource management}
	
	We use an \emph{orchestrator} in our experiments; an agent that monitors and manages resources while maintaining determinism.
	Each container needs three main resources: computing power to execute algorithms, memory to store and manipulate data, and I/O to interact with the environment. 
	The \emph{orchestrator}'s scheduling strategy maximizes resource utilization without going over the resource limits (over-subscription).
	Careful resource management allows efficient resource allocation while mitigating the risk of missing deadlines. 
	To ensure determinism, one can set the Worst Case Execution Time (WCET) of periodic deadline scheduled tasks to the maximum measured or expected amount, accepting to potentially under-utilize available resources. 
	We have conceived a solution that enables higher sharing, and thus saving, of low-use time-slices while accepting a low probability of deadline misses.
	To achieve this, the orchestrator confines the containers and running processes in their execution context, and monitors them to compute the probability of missed deadlines.
	If these probabilities exceed application specific thresholds, the orchestrator reschedules the respective tasks next in line onto a different, less critical resource.
	Thus, our orchestrator optimizes utilization for efficient resource allocation, while moderating the ensuing risks.
	
	Optimal resource management in full generality is hard to achieve. 
	To ease process allocation and resource assignment, we impose two initial assumptions.
	First, we assume independence of the real-time software running in containers for which we have neither specifications nor source code.
	Accordingly, the scheduling algorithm virtually simplifies to multiple single-core units where allocation reaches theoretical utilization rates of $69\%$ for Rate Monotonic (RM) scheduling and up to $100\%$ for Earliest Deadline First (EDF)~\cite{Buttazzo2011}.
	Second, we assume knowledge of all data for all possible real-time tasks.
	By experimental measurement and function fitting we retrieve a task's probability distribution of execution times and its WCET.
	Reducing the reserved CPU-budget to values close to the average task run-time increases the probability of exceeding new deadlines but opens part of the reserved slot to shared use.
	Equation \ref{eq:buffer} determines the size of this shared buffer.
	
	\begin{equation}
	\label{eq:buffer}
	Buffer = 1-\Sigma_n {\frac{n's\ CPU\ budget}{task\ n's\ period}}
	\end{equation}

	Next, we assume that the probability distribution function (PDF) of our tasks' run-time is normal.
	A task that executes a pure recurring calculation will always result in the same computation requirement.
	As the software does not contain any conditional branching, its effective run-time and thus the distribution's shape depend on system noise and I/O.
	We thus suppose, as a simplification, that the monitored tasks behave like a recurring computation with a normal-like distribution.
	This distribution is thus limited on the right bound by the tasks required computation time and shifted by varying latency, or I/O delays.
	According to this assumption, an experimentally acquired PDF of such a repetitive computation should closely match a normal distribution.
	Our experiments described in Section~\ref{sec:experiments and results} confirmed these assumptions.

	Starting with this assumption we can then compute the probability of a deadline miss of a group of tasks running on the same resources.
	The joint value for normally distributed execution times is given by
	\begin{equation*}
	P = \int_{U_{max}}^{\infty} N(\mu_{sum},\sigma_{sum}),
	\end{equation*}
	resulting in another normal distribution. $\mu_{sum}$ and $\sigma_{sum}$ are the arithmetic sums of mean and variance of the resulting probability distribution of each task.
	The probability of deadline miss is given by the area under the function, left of the maximum utilization factor $U_{max}$ (1 for pure EDF scheduling).
	With this probability the orchestrator then determines the need for reallocation.
	However, typical real-time tasks do not follow this assumption. 
	To reduce run-time variance and consequently diminish re-allocations, a target control software could further be optimized for quasi-static execution~\cite{Liu2005}.
	This reduces software branching and the resulting execution time variations, bringing the PDF closer to a simpler, normal distribution.
	Furthermore, in a second iteration, the implementation of the orchestration software will estimate and fit each tasks PDF through active run-time monitoring to typical distribution functions.

	\begin{figure}[tb]
		\centering
		\includegraphics[width=0.9\linewidth]{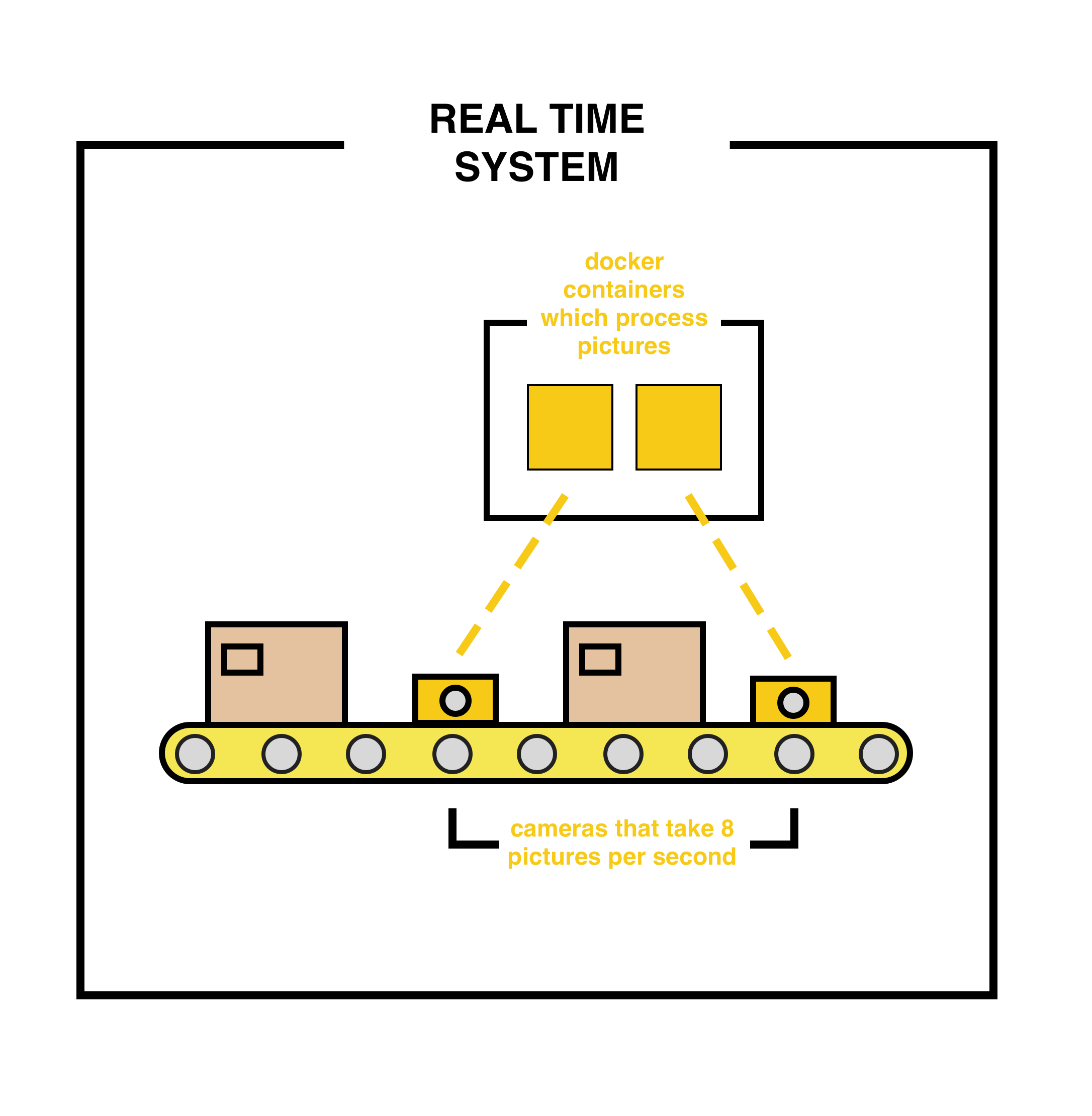}
		\caption{Example application of a manufacturing use case involving a conveyor belt.}
		\label{fig:belt}
	\end{figure}	

\begin{figure}[tb]
		\centering
		\includegraphics[width=\linewidth]{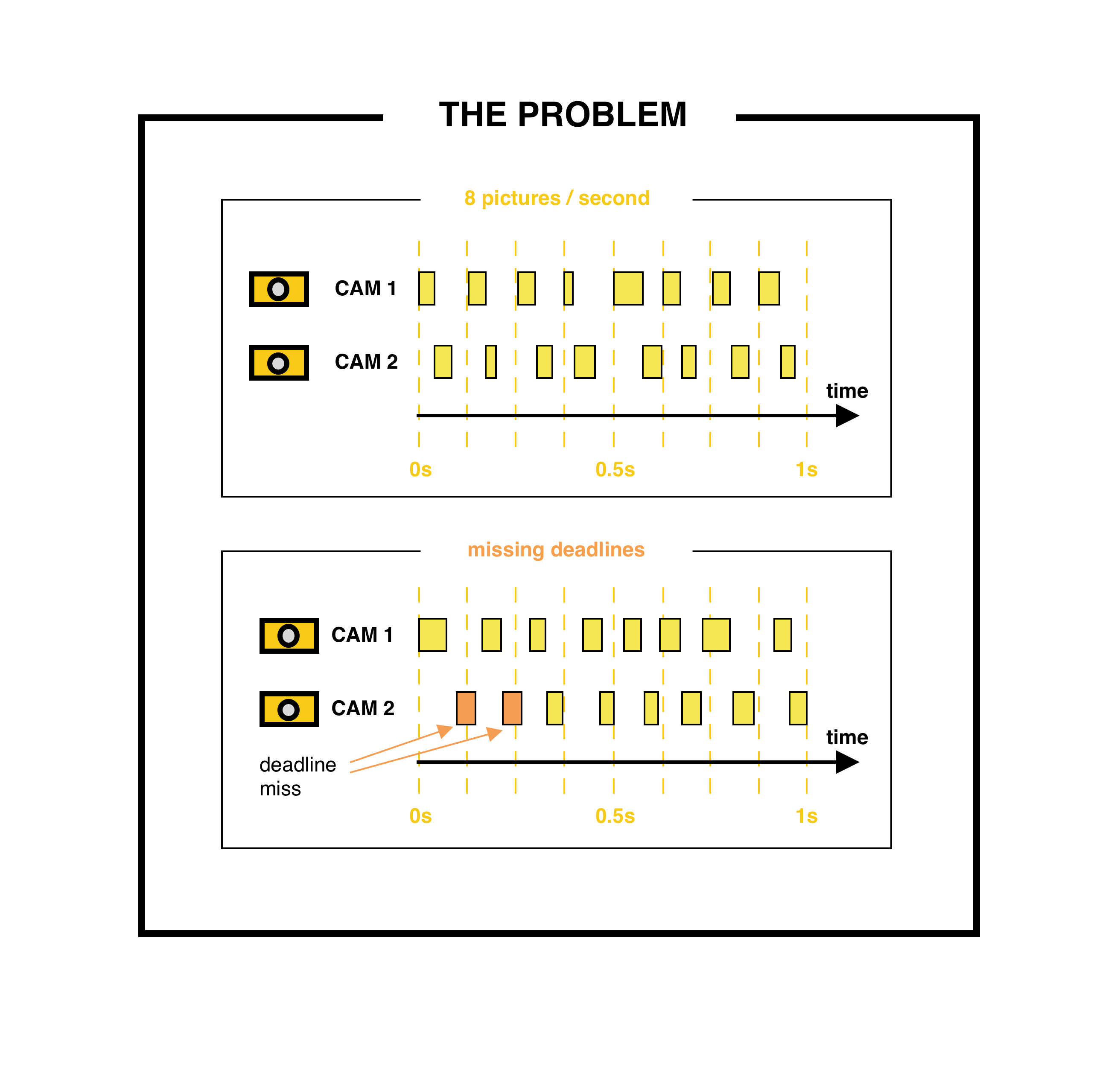}
		\caption{Image processing allocation, ideal vs deviation. Deadline misses may occur if images change significantly.}
		\label{fig:schedule}
	\end{figure}

	Figure \ref{fig:belt} shows an example for such a use case.
	Two cameras mounted along a conveyor belt take 8 pictures per second on both sides of a product, requiring an assessment (pass/fail) every $125ms$.
	If the image processing time of each picture remains below 50\% of the sampling time, both cameras can run on the same computing resources, saving one CPU, Figure~\ref{fig:schedule}.
	However, image processing times depend on the image structure. 
	A perfect product will always yield the same processing time. 
	Defects and lighting variations can however increase this value and thus cause missed deadlines, shown in the lower part of the figure.

	In such a situation, typical systems would need to reserve separate resources exclusively for the two cameras, resulting in resource waste.
	With traditional statically assigned resources, tasks use the remainder of a slot between average and peak value only when images differ from the ideal.
	To keep the failure probability low, the systems have to account for the expected WCET, reserving more resources than necessary most of the time.
	The orchestrator allows reclaiming part of this slot, increasing resource utilization, at a cost of potential deadline misses with probability $P \geq 0$.
	This permits the system to share the exceeding resources to other processes and react only if needed.

\begin{figure}[tb]
		\centering
		\includegraphics[width=\linewidth]{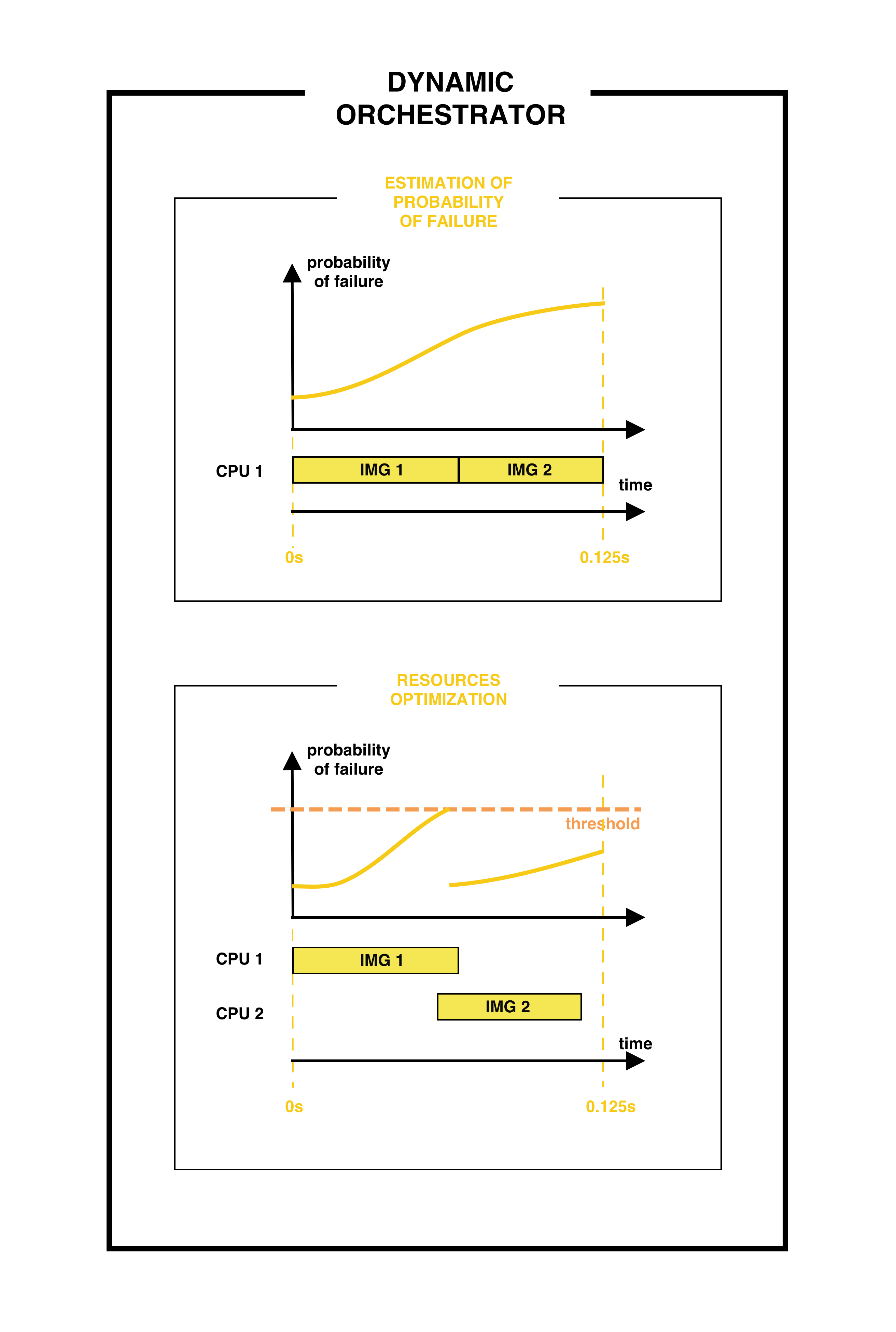}
		\caption{Image processing reallocation. The orchestrator tries to avoid a deadline miss by reallocating resources to other, lower critical areas.}
		\label{fig:reschedule}
	\end{figure}

	Figure~\ref{fig:reschedule}, shows an improved resource allocation for the same application scenario.
	Instead of locking both CPUs for image processing, the orchestrator can allocate both tasks to the same CPU as long as the failure risk remains low.
	The second CPU can be reused for other best effort tasks until the risk of failure exceeds the preset threshold, Figure~\ref{fig:reschedule} below.
	If this happens, the orchestrator reclaims the CPU and transfers tasks at risk of deadline miss to a now spare CPU.
	Estimating the probability values through real-time monitoring, the orchestrator can thus halve the required resources in this use-case for most of its run time.

	\section{Experiments and results}
	\label{sec:experiments and results}

	\begin{figure}[tb]
		\centering
		\includegraphics[width=\linewidth]{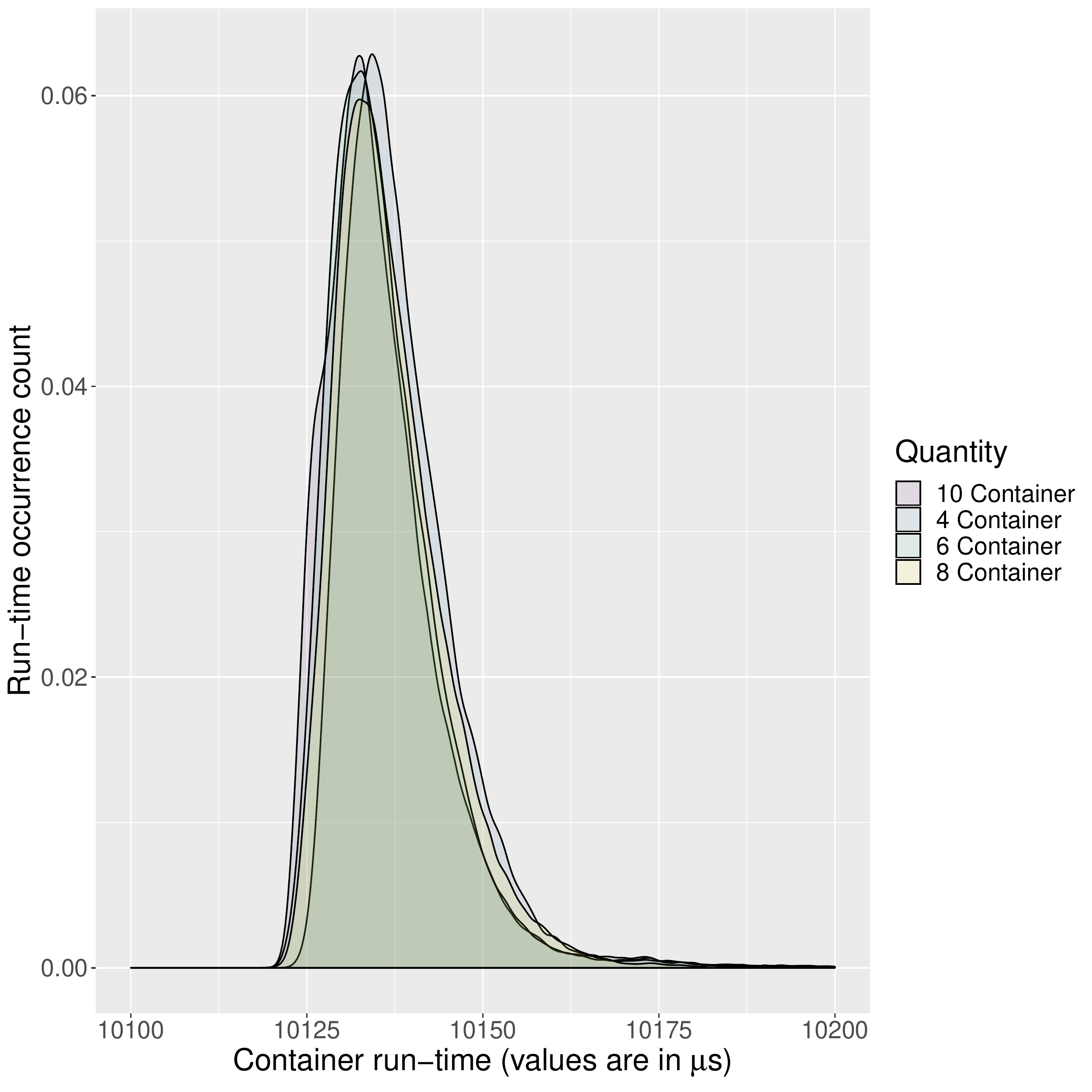}
		\caption{Experiment results for an up-to 10 container static resource allocation with $100ms$ period and $10ms$ run time on an {AWS} {C5.XLarge} instance with PREEMPT\_RT patch. The count is relative per container. Plotted from data sets in~\cite{Hoferetal2020-2}.}
		\label{fig:plot10}
	\end{figure}
	
	In recent experiments we tested the viability of an off-the-shelf approach in different phases.
	Firstly, we picked containerization tools and operating systems, and verified ease of use and maintainability against multiple criteria.
	Then, we tested their latency and deterministic performance in four system setups~\cite{Hoferetal2019}.	
	Secondly, we extended the experiments by adding an orchestration tool and static scheduled tasks. We observed then the tasks' behavior in changing load environments to determine the system's run time determinism~\cite{Hoferetal2020-2}.
	Lastly, with new experiments we will explore the mutation of latency and determinism when dynamic resource migration and additional I/O come into play.

	In our first set of experiments we assessed the feasibility of real-time task execution with COTS products and configurations in a virtualized environment~\cite{Hoferetal2019}.
	In the experiments, we targeted systems under stress and analyzed task latency across different configurations.
	We then proposed four solutions suitable for migration to application containerization while maintaining determinism.
	All four systems run our favorite configuration pick as their underlying operating system, i.e. Ubuntu LTS distributions with the PREEMPT\_RT patch.
	As expected, bare-metal solutions ensure the most deterministic behavior for hard real-time requirements on shared resources. 
	While virtualized instances performed similarly good in some occasions, virtual machines with dedicated computing resources are to be preferred.
	They outperformed, on average, latency and deterministic behavior of the more economic and generic computing instances.
	As the latter do not ensure resource exclusiveness, they might serve as compromise between performance and cost.
	In conclusion, the promising results confirmed the feasibility of a migration to IAAS solutions.

	\begin{figure}[tb]
		\centering
		\includegraphics[width=\linewidth]{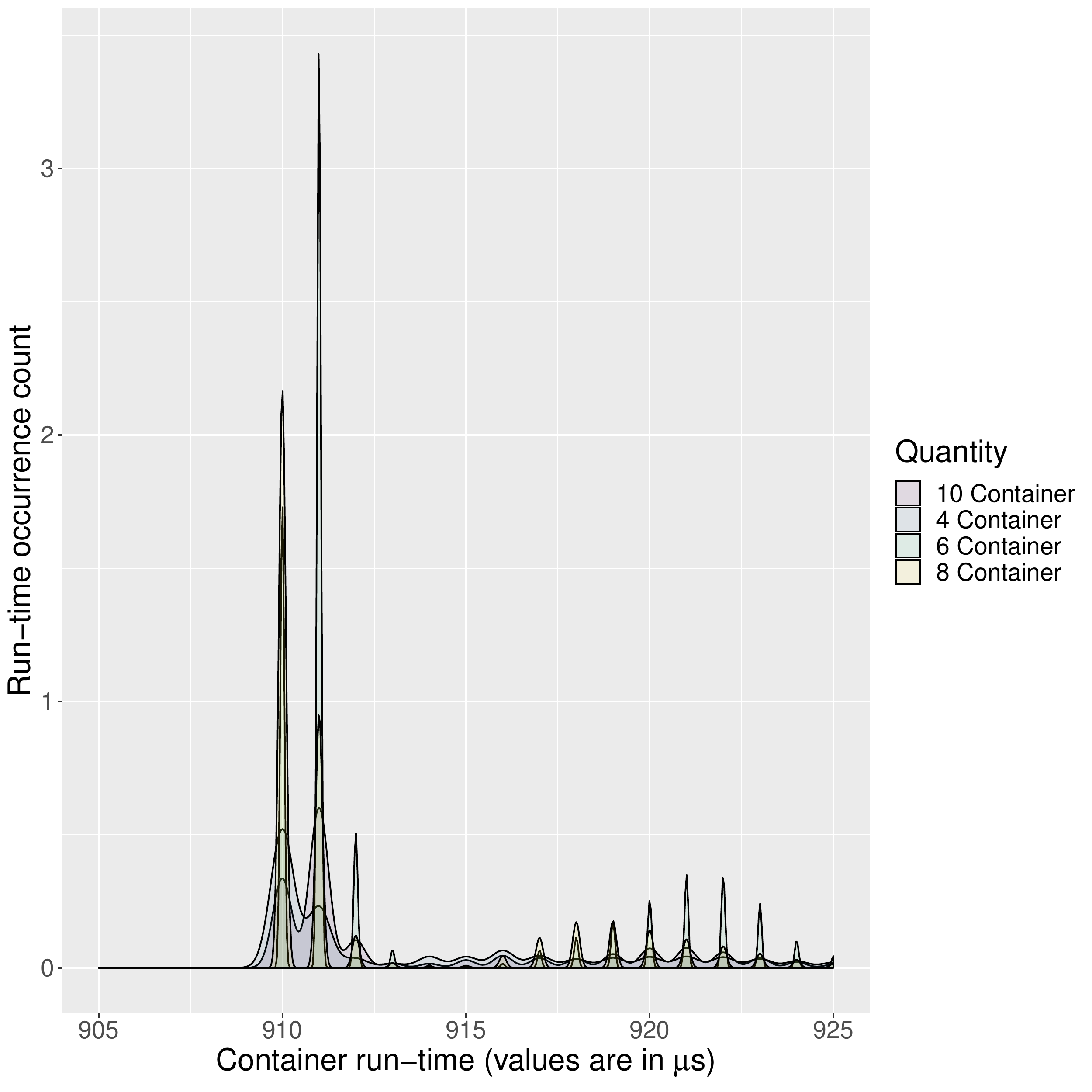}
		\caption{Experiment results for the same set-up of Figure~\ref{fig:plot10}, with $10ms$ period and $0.9ms$ run time. Source:~\cite{Hoferetal2020-2}.}
		\label{fig:plot1}
	\end{figure}
	{
	\renewcommand{\arraystretch}{1.5}
	\begin{table*}[tb] 
		\centering
		\caption{Summary for the experiments of Figure~\ref{fig:plot10}.
			Values are in $\mu s$ and show all containers averages (AVG), an asterisk notes a missed deadline; absolute distances from averages, skew (SKW); and the standard deviation of the container with the highest skew (SD\_MX).}
		\begin{tabular}{l *9{c}}
			\toprule
			\textbf{Configuration of containers} & \multicolumn{3}{c}{\bfseries Bare Metal}& \multicolumn{3}{c}{\bfseries Aws T3.Xlarge}& \multicolumn{3}{c}{\bfseries Aws C5.Xlarge}\\
			instance number and relative system load & \textbf{AVG} & \textbf{SKW} & \textbf{SD\_MX}& \textbf{AVG} & \textbf{SKW} & \textbf{SD\_MX}& \textbf{AVG} & \textbf{SKW} & \textbf{SD\_MX}\\
			\hline
			4 units, system load < 50\% & 10712 & 0/8 & 31.78 & 10072 & 14/16 & 45.69 & 10139 & 2/2 & 12.03\\
			5 units, system load < 60\%  & 10614 & 8/10 & 35.66 & 10056 & 14/16 & 35.46 & 10310 & 1/3 & 68.98\\
			6 units, system load < 70\%  & 10132 & 3/10 & 34.87 & 10038 & 4/7 & 23.30 & 10136 & 1/2 & 10.55\\
			7 units, system load < 80\% & 10115 & 0/11 & 31.25 & 10166 & 49/57 & 123.36 & 10138 & 1/2 & 11.08\\
			8 units, system load < 90\% & 10104 & 3/12 & 32.20 & 10052 & 10/16 & 41.55 & 10137 & 1/2 & 9.04\\
			9 units, system load < 100\% & 10356 & 3/8 & 29.83 & 10059 & 5/16 & 126.17 & 10138 & 1/2 & 9.94\\
			10 units, system load $\approx$100\% & 10089* & 4/10 & 46.29 & 10027* & 1/3 &  12.75 & 10136* & 1/2 & 9.91\\
			\bottomrule
		\end{tabular}
		\label{tab:resgrp4}
	\end{table*}
	}

	In a follow-up experiment we analyzed static real-time task grouping and scheduling to optimize resource allocation~\cite{Hoferetal2020-2}.
	We used our orchestration tool to manage schedule, environment and resource settings.
	In the same four environments the experiments showed that, for computation only, the subscription of resources can exceed 90\% without deadline misses. 
	Figure~\ref{fig:plot10} shows the run time distributions of an experiment with increasing container count, while in Table~\ref{tab:resgrp4} we summarize numerical results.
	According to our assumptions in Section~\ref{sec:resource management}, each container runs a single task with a recurring calculation. 
	As expected, all the plots present a slightly skewed normal distribution with a left-side cut-off.
	Consequently, system latency, interrupts and resulting operating system overhead are the main factors responsible for this skew.
	Furthermore, with a computation time of $10ms$ and period of $100ms$, and accounting for the operating system overhead, the ten container setup oversubscribes the CPU. 
	Nonetheless, the distribution displays the same normal shape, confirming the approaches' stability.
	These results confirm the assumption that a normal PDF, as presented in Section~\ref{sec:resource management}, may get good probability estimation results.
	With shorter period and computation time, i.e., $0.9ms$ and $10ms$ respectively, the mentioned system overhead gets more visible.
	In Figure~\ref{fig:plot1}, we see this configuration where blocking I/O calls and higher priority interrupts notably influence the execution queue.
	Due to the short run-time, the delays induced by system noise are greater than the tasks original PDF and create duplicate skewed normal distributions.
	Although these deviations, the base shape of the histogram results still normal.
	How the plot changes with programmed I/O and consequent interrupts will be investigated in upcoming experiments.
		
	In our next step we investigate the active resource management and reallocation strategies.
	In these test, the orchestrator schedules and reallocates tasks on demand according to their failure probability.
	We test the performance and stability of our approach for resource management through two case studies.
	The first explores the behavior of our approach in varying load conditions, analyzing a group of event driven real-time tasks, called workers, which are fed with information on variable intervals. 
	The second study helps to assess the orchestration behavior and effectiveness in more complex applications with strict real-time requirements.
	This includes mixed real-time applications, event driven and polling-based, that have different execution periods and real-time behavior.
	We perform our monitoring and reallocation tests initially using a naive algorithm.
	In this simple setting, the orchestrator reallocates tasks to the resource that give the best match on its run-time  parameters, e.g., same period or computation ratio.
	Once we established a baseline, we gradually optimize the code with statistic allocation methods such as the Monte Carlo bin problem~\cite{HarrisAltiparmak2019}.
	We expect to see the amount of I/O performed by the system to be the major influencing factor for resource reuse and re-schedulability.
	How much these effects account for and possible orchestration consequences in an industrial real study case are hard to predict and open for future study.
	
	A conclusive effort will be the analysis of security issues that such a migration entails.
	In a four phase program~\cite{Hofer2018-2} we analyze issues that arise in the different components composing new Industry 4.0 architectures, including virtual and traditional systems.
	Through a cross-domain layer based technique, first presented in Hofer and Russo~\cite{HoferRusso2020}, we identify vulnerabilities based on known attack schemes. 
	The resulting off-line security analysis summarizes thus expected security issues to address. 
	
	\section{Conclusions}
	Using dynamic, probabilistic resource allocation and scheduling, our orchestration approach addresses the demand for economic real-time capable computing. 
	Mixed criticality applications prepare environments for shared execution of tasks with different priorities, enabling us to exploit this variety to bridge peak demands for hard real-time tasks.
	The resulting resource savings reduce installation, operation and maintenance costs of smart industrial plants, leveraging virtualization technologies for their control applications.
	However, unpredictability of I/O and operating system fluctuations make it hard to estimate the variability and impact in real use cases. 
	
	Future work will explore application to industrial automation use cases, obtainable savings and the difference in impact between synthetic and real use cases. 
	In new experiments we will explore the efficiency, estimation and performance of the orchestration with tasks that present a non-normal distributed probability function.
	We will further explore additional orchestration algorithms for non-deadline scheduled tasks and their efficiency and investigate possible security issues that might arise through control migration. 
	
	\bibliographystyle{IEEEtran}
	\bibliography{papers.bib}

\end{document}